\documentclass{article}

\usepackage[preprint]{main}

\usepackage[utf8]{inputenc} 
\usepackage[T1]{fontenc}    
\usepackage{hyperref}       
\usepackage{url}            
\usepackage{booktabs}       
\usepackage{amsfonts}       
\usepackage{amsmath} 
\usepackage{nicefrac}       
\usepackage{microtype}      
\usepackage{xcolor}         
\usepackage{tabularx}       
\usepackage{array}
\usepackage{graphicx}
\usepackage{algorithm} 
\usepackage{algpseudocode}
\usepackage{subfiles}

\title{Homogenized \textit{C. elegans} Neural Activity and Connectivity Data}

\author{%
  Quilee Simeon\thanks{Corresponding author.} \\
  MIT \\
  \texttt{qsimeon@mit.edu} \\
  \And
  Hector Astrom \\
  MIT \\
  \texttt{hastrom@mit.edu} \\
  \And
  Anshul Kashyap \\
  UC Berkeley \\
  \texttt{anshulkashyap@berkeley.edu} \\
  \And
  Konrad P. Kording \\
  University of Pennsylvania \\
  \texttt{kording@upenn.edu} \\
  \And
  Edward S. Boyden\thanks{Senior author.} \\
  HHMI, MIT \\
  \texttt{edboyden@mit.edu} \\
}

\begin{document}

\maketitle

\begin{abstract}
There is renewed interest in modeling and understanding the nervous system of the nematode \textit{Caenorhabditis elegans} (\textit{C. elegans}), as this small model system provides a path to bridge the gap between nervous system structure (connectivity) and function (physiology). However, existing physiology datasets, whether involving passive recording or stimulation, are in distinct formats, and connectome datasets require preprocessing before analysis can commence. Here we compile and homogenize datasets of neural activity and connectivity. Our neural activity dataset is derived from 12 \textit{C. elegans} neuroimaging experiments, while our connectivity dataset is compiled from 9 connectome annotations based on 3 primary electron microscopy studies and 1 signal propagation study. Physiology datasets, collected under varying protocols, measure calcium fluorescence in labeled subsets of the worm’s 300 neurons. Our preprocessing pipeline standardizes these datasets by consistently ordering labeled neurons and resampling traces to a common sampling rate, yielding recordings from approximately 900 worms and 250 uniquely labeled neurons. The connectome datasets, collected from electron microscopy reconstructions, represent the entire nervous system as a graph of connections. Our collection is accessible on HuggingFace, facilitating analysis of the structure-function relationship in biology using modern neural network architectures and enabling cross-lab and cross-animal comparisons.
\end{abstract}

\break

\section{Introduction}

Understanding neural dynamics and their relationship to underlying structure remains one of the key challenges in neuroscience. Recent advances in computational modeling and data acquisition have reinvigorated interest in the neural system of the nematode \textit{Caenorhabditis elegans} (\textit{C. elegans}), a model organism uniquely suited for bridging the gap between neural connectivity and function. With its fully mapped synaptic connectome \cite{white1986structure, witvliet2021connectomes}, consisting of approximately 300 neurons \cite{cook2019connectomes}, and its extensively studied neuronal dynamics \cite{goodman1998activecurrents, yemini2021neuropal}, \textit{C. elegans} provides an ideal platform for investigating how neural structure constrains or influences physiology. 

Recent research has focused on leveraging the \textit{C. elegans} connectome for generative models of neural development \cite{richter2024smallbrain} and linking whole-brain activity to synaptic connectivity \cite{creamer2024connectomeactivity}. These studies exemplify its potential for elucidating how neural signaling is determined or constrained by connectivity and for inspiring biologically grounded computational frameworks \cite{hasani2022continuous}. Furthermore, the transparent body of \textit{C. elegans}, combined with its well-characterized neural architecture, makes it an excellent candidate for whole-brain functional imaging using calcium fluorescence sensors \cite{yemini2021neuropal}. Calcium imaging enables the capture of neural dynamics at single-neuron resolution across many neurons simultaneously \cite{chen2013ultrasensitive}, serving as a proxy for neuronal activity. Although calcium fluctuations may primarily reflect local processes in neurites rather than whole-neuron computations \cite{grienberger2012imaging}, compact neurons like those in \textit{C. elegans} likely rely on such localized signaling mechanisms, making this an informative approach.

The combination of calcium imaging datasets with the full connectome, mapped via electron microscopy \cite{white1986structure, cook2019connectomes, witvliet2021connectomes}, offers an unprecedented opportunity to investigate how connectivity influences neural function. Unified datasets of structure and function can provide valuable insights into neural activity propagation, informing both the development of artificial intelligence (AI)-based neural simulations and the understanding of mechanisms that generalize to more complex organisms \cite{haspel2023reverseengineer}.

However, calcium imaging datasets from \textit{C. elegans} experiments vary widely in their experimental conditions, labeled neurons, and sampling rates, complicating cross-study comparisons and unified modeling efforts. Similarly, connectome datasets—whether physical or functional—require significant preprocessing to facilitate analysis or computational modeling.

To address these challenges, we have created two integrated datasets. The first aggregates neural dynamics from 12 calcium imaging studies, and the second compiles connectome data from three electron microscopy wiring studies \cite{white1986structure, cook2019connectomes, witvliet2021connectomes} and one functional signal propagation study \cite{randi2023neural}. Together, these datasets combine structure and function, offering an invaluable resource for computational modelers aiming to construct biologically grounded neural network models. 

Our preprocessing pipeline standardizes neural activity data through normalization, resampling to a common rate, and consistent organization into data structures. Neural recordings include data from approximately 900 worms and 250 uniquely labeled neurons. The connectome data is represented as a graph with attributes for neuron positions, synaptic connections, and additional metadata, facilitating downstream analyses that integrate structure and function. 

These datasets are available as open-source resources on the HuggingFace platform, ensuring accessibility for researchers across neuroscience and machine learning. By unifying and releasing these datasets, we aim to catalyze the development of computational models that capture the structure-function relationships in small neural systems, enabling robust cross-lab and cross-animal comparisons.

\section{Methods}

\subsection{Code Repository}
We provide an open-source \href{https://github.com/qsimeon/worm-data-preprocess}{code repository } that implements the entire preprocessing pipeline we describe in following sections. Our preprocessing is combination of utilities (classes, functions, files and links) that facilitate the extraction and processing of a collection of \textit{C. elegans} neural and connectome data into standardized formats. 

The main classes in the repository are \texttt{NeuralBasePreprocessor} and \texttt{ConnectomeBasePreprocessor}. The former manages processing of neural activity data as measured by calcium fluorescence imaging \cite{tian2009imaging}. The latter manages the processing of synaptic connectivity data as measured by anatomical counts from electron microscopy (EM) \cite{white1986structure}.

The main functions in the repository are \texttt{pickle\_neural\_data()} and \texttt{preprocess\_connectome()}. The \texttt{pickle\_neural\_data()} function extracts and processes raw calcium fluorescence data from each of our 12 neural activity source datasets, standardizing them into neural activity matrices that are saved as compressed files for efficient storage and retrieval. The \texttt{preprocess\_connectome()} function processes raw connectome data from source files into graph tensors, which are formatted for compatibility with graph neural network (GNN) frameworks such as PyTorch Geometric \cite{fey2019torchgeometric}. This preprocessing step facilitates downstream tasks like connectivity-informed modeling of neural activity. The 10 distinct connectome source files are detailed in Table \ref{table:open_source_connectome_datasets}.

Our code infrastructure allows us to process raw neural and connectome data from different source datasets in a consistent and standard way.

\begin{figure}[!htbp]
\centering
\includegraphics[width=1.0\textwidth]{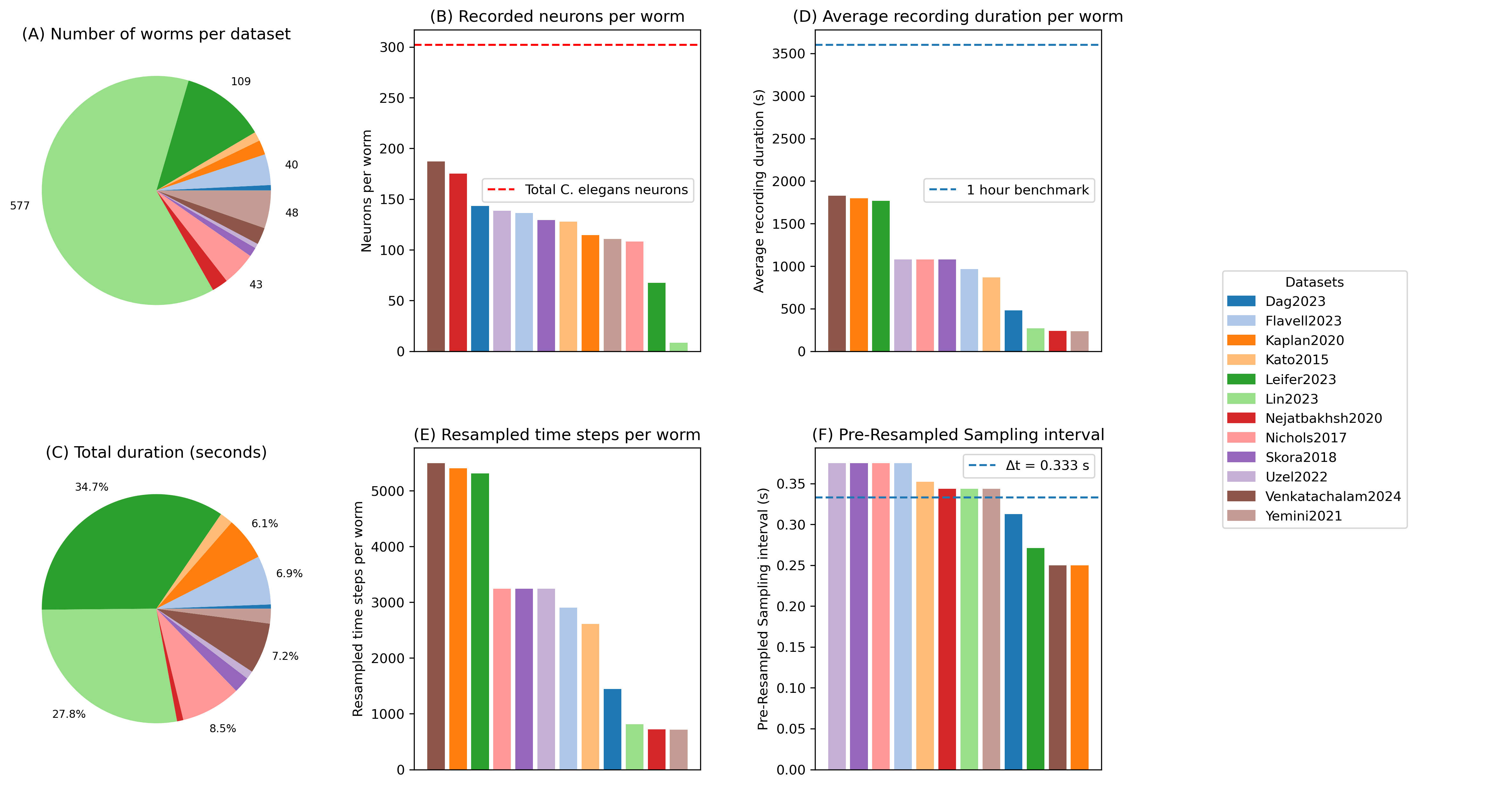}
\caption{\textbf{Overview of Neural Activity Datasets.} (A) Distribution of the number of worms in each experimental dataset. (B) Number of recorded neurons per worm compared to the total neuronal population of \textit{C. elegans}. (C) Total duration of recorded neural activity for each dataset. (D) Average recording duration per worm, with one hour of calcium imaging as a benchmark. (E) Number of resampled time steps per worm, and (F) Pre-resampled sampling intervals for recorded neural activity. The horizontal dashed line in (F) indicates the target resampled time step ($\Delta t = 0.333$ seconds) used in our preprocessing pipeline.}
\label{fig:neural_activity_overview}
\end{figure}

\subsection{Calcium Fluorescence Data}
\begin{figure}[!htbp]
\centering
\normalsize
\includegraphics[width=\textwidth]{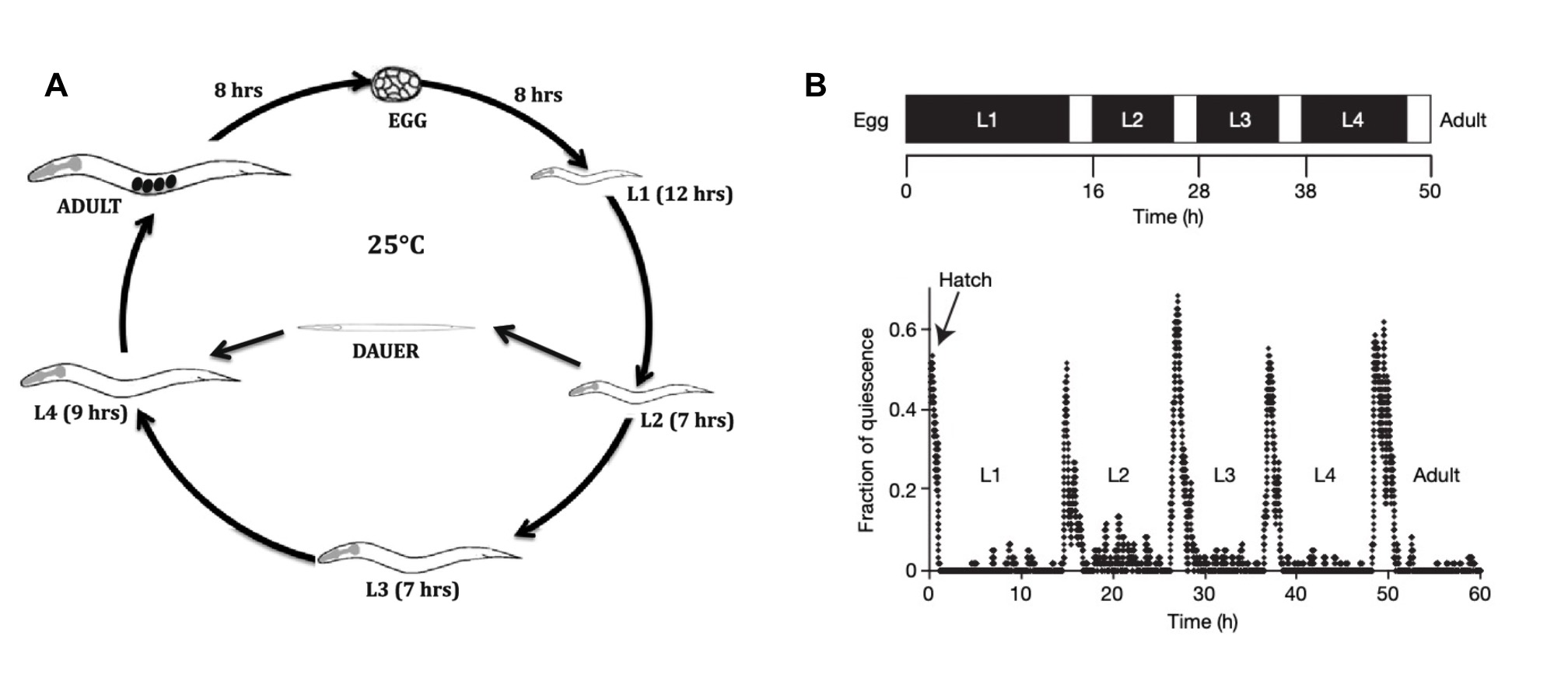}
Appendix Figure 2: \textbf{Developmental stages and quiescence in \textit{C. elegans}.} 
(A) Diagram of the developmental stages of \textit{C. elegans}, progressing from egg to adult, showing the timing of each larval stage at 25$^\circ$C. Adapted from \cite{pandey2014unc53}. 
(B) Behavioral quiescence during lethargus periods, which occur at the transitions between larval stages. Adapted from Figure 1 of \cite{raizen2008lethargus}.
\label{fig:celegans_developmental_stages}
\end{figure}

We processed neural activity datasets from 12 open-source studies of \textit{C. elegans}, each measuring calcium fluorescence ($\Delta F / F_0$) in subsets of the worm’s neurons under various conditions \cite{randi2023neural, atanas2023brainwide, uzel2022hubneurons, yemini2021neuropal, kaplan2020nested, skora2018energy, nichols2017brain, kato2015global, lin2023functional, dag2023serotonergic}. These datasets include a variety of experimental protocols, such as freely moving, immobilized, and optogenetically stimulated animals. The number of worms and identified neurons varied across datasets as shown in Table \ref{table:open_source_neural_datasets}. All worms were hermaphrodites at developmental stages no earlier than L4 \cite{raizen2008lethargus}, with most being in adulthood (Figure \ref{fig:celegans_developmental_stages}). 

\begin{table}[!htbp]
\caption{\textbf{Calcium Fluorescence Neural Activity Datasets Metadata.} Metadata for calcium fluorescence neural activity datasets collected from various sources. The table includes the number of worms, the average number of labeled and recorded neurons, the range of labeled neurons observed, and summary experimental protocol details.}
\centering
\scriptsize
\renewcommand{\arraystretch}{1.3} 
\begin{tabularx}{\textwidth}{|c|c|c|l|l|X|}
\hline
\textbf{Index} & \textbf{Dataset Name} & \textbf{Num. Worms} & 
\textbf{\begin{tabular}[c]{@{}l@{}}Mean Num. \\ Neurons \\ (labeled, recorded)\end{tabular}} & 
\textbf{\begin{tabular}[c]{@{}l@{}}Num. Labeled \\ Neurons \\ (min, max)\end{tabular}} & 
\textbf{Experimental Protocol} \\
\hline
1 & Kato2015 \cite{kato2015global} & 12 & (42, 127) & (31, 51) & Wild-type; L4 stage; immobilized in a microfluidic device \\
2 & Nichols2017 \cite{nichols2017brain} & 44 & (34, 108) & (23, 43) & Wild-type and \textit{npr-1} mutants; Lethargus/pre-lethargus stages; immobilized \\
3 & Skora2018 \cite{skora2018energy} & 12 & (46, 129) & (39, 55) & Wild-type; fasting/starvation conditions; immobilized \\
4 & Kaplan2020 \cite{kaplan2020nested} & 19 & (36, 114) & (23, 51) & Wild-type; L4 stage; freely moving and immobilized \\
5 & Nejatbakhsh2020 \cite{nejatbakhsh2020} & 21 & (173, 175) & (163, 184) & Transgenic (GCaMP6); L4 to adult stages; during chemotaxis/decision-making tasks; immobilized \\
6 & Yemini2021 \cite{yemini2021neuropal} & 49 & (110, 110) & (33, 179) & Transgenic (OH15500); adult stage; immobilized \\
7 & Uzel2022 \cite{uzel2022hubneurons} & 6 & (50, 138) & (46, 58) & Wild-type; adult stage; Well-fed but immobilized \\
8 & Dag2023 \cite{dag2023serotonergic} & 7 & (100, 143) & (87, 110) & Wild-type; adult stage; freely moving with subsequent NeuroPAL immobilization \\
9 & Leifer2023 \cite{randi2023neural} & 110 & (63, 67) & (13, 98) & Wild-type; adult stage; immobilized \\
10 & Lin2023 \cite{lin2023functional} & 577 & (8, 8) & (1, 22) & Wild-type; adult stage; immobilized \\
11 & Flavell2023 \cite{flavell2023} & 40 & (88, 136) & (64, 115) & Transgenic (GCaMP7f, mNeptune2.5); adult stage; freely moving active and quiescent states \\
12 & Venkatachalam2024 \cite{venkatachalam2024website} & 22 & (187, 187) & (185, 189) & Wild-type; adult stage; immobilized in microfluidic chips \\
\hline
\end{tabularx}
\label{table:open_source_neural_datasets}
\end{table}

\subsubsection{Overview of Neural Activity Datasets}
We analyzed neural activity datasets collected from 12 independent studies, which varied in terms of the number of worms recorded, the number of neurons labeled, the duration of recordings, and sampling intervals. Figure \ref{fig:neural_activity_overview} provides some summary statistics on the neural activity data collected from each of these datasets.

\subsubsection{Brief Note on Notation}
In this section, we define the key mathematical notation used throughout our preprocessing pipeline and neural activity data representation. This note aims to provide clarity and consistency for readers, particularly when interpreting symbols used in equations and methods. For a comprehensive description of the symbols, including those specific to other parts of the pipeline, refer to Appendix Table \ref{table:neural_notation_table}.

We represent the neural activity data for a worm \( k \) as a time-series matrix \( \mathbf{X}^{(k)} \in \mathbb{R}^{T_k \times D} \), where \( T_k \) is the number of recorded time points, and \( D = 300 \) is the total number of canonical neurons in \textit{C. elegans}. This excludes the canal-associated neurons (CAN), which are not typically included in functional studies \cite{stefanakis2015panneuronal, ripoll2023neuropeptidergic}. Each row of \( \mathbf{X}^{(k)} \) corresponds to a snapshot of neural activity across all \( D \) neurons at a single time point, while each column corresponds to the activity trace for a specific neuron over time.

Key conventions include:
\begin{itemize}
    \item \textbf{Activity Snapshot}: The neural activity at time \( t \) is denoted \( \mathbf{X}^{(k)}[t] \in \mathbb{R}^D \), representing the activity of all \( D \) neurons at that moment.
    \item \textbf{Indexing}: We use \( 0 \)-based indexing for time points (e.g., \( t = 0, 1, \dots, T_k - 1 \)) and \( 1 \)-based indexing for neurons (e.g., \( i = 1, 2, \dots, D \)).
    \item \textbf{Vector Orientation}: All vectors (e.g., \( \mathbf{X}^{(k)}[t] \)) are column vectors unless explicitly stated otherwise.
\end{itemize}

This notation provides the foundation for the mathematical operations described in the neural data processing pipeline. For additional details on notation, see Appendix Table \ref{table:neural_notation_table}.

\subsubsection{Neural Data Processing} \label{neural_data_processing}
The entire pipeline is implemented in the \texttt{NeuralBasePreprocessor} class, which handles dataset-specific preprocessing tasks, including file format loading, trace extraction, and metadata creation.

\begin{figure}[!htbp]

\centering

\includegraphics[width=1.0\textwidth]{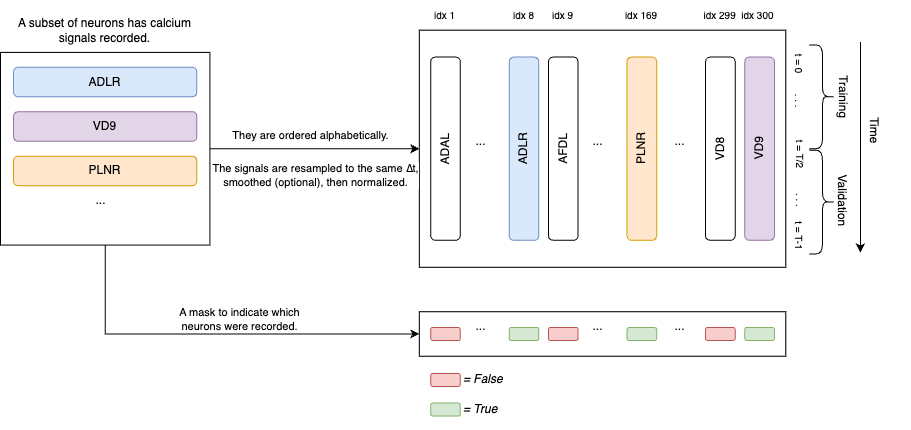}

\caption{\textbf{Processed Neural Activity Data Structure.} Illustration of the structure of our neural activity data as a time-series matrix, where each column represents a neuron, ordered alphabetically by their canonical names in \textit{C. elegans}. The calcium signals are resampled, (optionally) smoothed, and normalized to the same rate ($1/\Delta t$ Hz). A binary mask is generated to indicate which neurons were labeled for each worm. }

\label{fig:neural_pipeline}

\end{figure}

The neural activity data is preprocessed using a custom pipeline (Figure \ref{fig:neural_pipeline}) which follows the general outline given in Algorithm 1 and described in more detail next.

a). \textbf{Data Download and Extraction.} The raw data from each dataset is downloaded from its respective source and extracted. The function \texttt{download\_url()} downloads the dataset, and \texttt{extract\_zip()} decompresses the data into a local directory while preserving the original folder structure.

b). \textbf{Custom Dataset Implementations.} Each source dataset has a custom class that inherits from the \texttt{NeuralBasePreprocessor} parent class, facilitating data extraction for each unique format. The preprocessor classes ensure that each dataset is loaded correctly, followed by the main preprocessing steps in order: resampling, optional smoothing, and normalization. Adding a new source dataset to the pipeline is as easy as defining a new child preprocessor class that implements any loading and processing logic that needs to be customized to that particular source; for example, if the new source's data is stored in a special file type.

c). \textbf{Resampling.} The first preprocessing step resamples the raw data to a fixed time interval $\Delta t$, using linear interpolation for datasets with different temporal resolutions. The new time index $t^{\prime}$ is calculated by:

\[
t^{\prime}=\left\lfloor\frac{t \cdot \Delta t_{\text{original}}}{\Delta t}\right\rfloor
\]

The resampled traces $\mathbf{X}_{\text{resampled}}^{(k)}$ are computed from the raw data $\mathbf{X}^{(k)}_{\text{raw}}$ as:

\[
\mathbf{X}_{\text{resampled}}^{(k)}[t^{\prime}] = \mathbf{X}^{(k)}_{\text{raw}}[t] + \left(\mathbf{X}^{(k)}_{\text{raw}}[t+1] - \mathbf{X}^{(k)}_{\text{raw}}[t]\right) \cdot (t^{\prime} - t)
\]

This ensures consistent temporal resolution across datasets. We use $\Delta t \approx 0.333$ seconds as a compromise to retain high- and low-frequency dynamics. After resampling, we set:
\[
\mathbf{X}^{(k)} \leftarrow \mathbf{X}_{\text{resampled}}^{(k)}
\]

Resampling is done as the first preprocessing step to ensure meaningful \textit{cumulative} data, as we'll discuss later on in step 5 of preprocessing, normalization. 

d). \textbf{Optional Smoothing.} Following resampling, the calcium traces can optionally be smoothed using an exponentially weighted moving average (EWMA). For each neuron, we apply causal smoothing as:

\[
\mathbf{X}^{(k)}_{\text{smooth}}[t] = \alpha \mathbf{X}^{(k)}[t] + (1-\alpha) \mathbf{X}^{(k)}[t-1]
\]

where $\alpha$ is a smoothing hyperparameter set in \texttt{preprocess/config.py}. For our released neural dataset, smoothing was not applied ($\alpha = 1$) as prior preprocessing had already been performed. Calcium fluorescence data inherently reflects low frequency neural dynamics. Smoothing acts as a further low pass over the resampled data and may therefore be unnecessary.

e). \textbf{Neuron Trace Normalization.} As the final preprocessing step, the preprocessors normalize (or $z$-score) the neural activity across the neuron dimension. Let $\mathbf{X}^{(k)}_{\text{input}} := \mathbf{X}^{(k)}_{\text{smooth}}$ if smoothing was applied; otherwise $\mathbf{X}^{(k)}_{\text{input}} := \mathbf{X}^{(k)}_{\text{resampled}}$. For each neuron trace $i$ in worm $k$, we compute:

\begin{equation}
    \mathbf{X}^{(k)}_{\text{norm}} = \frac{\mathbf{X}^{(k)}_{\text{input}} - \boldsymbol{\mu}}{\boldsymbol{\sigma}},
\end{equation}

where $\boldsymbol{\mu}, \boldsymbol{\sigma} \in \mathbb{R}^{D}$ are column vectors of the mean and standard deviation, respectively, for each neuron, calculated across the temporal dimension.

This standard normalization method uses the full time-series to compute $\boldsymbol{\mu}$ and $\boldsymbol{\sigma}$, which introduces information from future states into earlier time points, thereby violating causality. While this may be acceptable for some applications, respecting causality could be crucial in certain cases, such as for real-time or streaming neural activity models.

f). \textbf{Masking and Subsetting.} We also create a binary mask for each worm, $\mathbf{M}^{(k)} \in \{0,1\}^D$, where $D = 300$ represents the total number of canonical neurons in \textit{C. elegans}. Each element $\mathbf{M}^{(k)}_i = 1$ indicates that neuron $i$ was labeled in worm $k$, while $\mathbf{M}^{(k)}_i = 0$ indicates that neuron $i$ was not labeled. 

In some of the source datasets we processed there were instances of calcium fluorescence measurements from neurons which were not identified with a label. For simplicity, we will not make a distinction in this work between unlabeled neurons and unmeasured neurons and may use the two terms interchangeably.

The mask $\mathbf{M}^{(k)}$ is stored alongside $\mathbf{X}^{(k)}$. We do not apply the mask to modify the neural data matrix (e.g. zeroing out measurements for unlabeled neurons) but instead retain it as a secondary data structure that can be used in downstream modeling and analysis (e.g., selecting only labeled neurons). Therefore, the pair $(\mathbf{X}^{(k)}, \mathbf{M}^{(k)})$ constitute the final output of the neural data processing half of our preprocessing pipeline. 

\begin{algorithm}
    \caption{Neural Data Preprocessing Pipeline}
    \begin{algorithmic}
        \State \textbf{Require} $\quad \alpha, \quad \Delta t, \quad \mathrm{all\_sources} := \{\text{Kato2015, ..., Venkatachalam2024}\}$
        
        \For{each dataset $\mathcal{D}_{\mathrm{raw, source}~i}$ in $\mathcal{D}_{\mathrm{raw, all\_sources}}$}
            \For{each worm $k$ in $\mathcal{D}_{\mathrm{raw, source} i}$}
                \State Extract calcium fluorescence traces from source files: $\mathbf{X}^{(k)}_{\text{raw}}$ 
                \State Normalize traces: $\mathbf{X}^{(k)}_{\text{norm}} \gets \text{normalize}(\mathbf{X}^{(k)}_{\text{raw}})$
                \If{Smooth traces}
                    \State $\mathbf{X}^{(k)}_{\text{original}} \gets \text{smooth}(\mathbf{X}^{(k)}_{\text{norm}}, \alpha)$
                \Else
                    \State $\mathbf{X}^{(k)}_{\text{original}} \gets \mathbf{X}^{(k)}_{\text{norm}}$
                \EndIf
                \State Resample traces: $\mathbf{X}^{(k)}_{\text{resample}} \gets \text{resample}(\mathbf{X}^{(k)}_{\text{original}}, \Delta t)$
                \
                \State Processed neural data: $\mathbf{X}^{(k)} \gets 
 \mathbf{X}^{(k)}_{\text{resample}}$
                \State Labeled neuron mask: $\mathbf{M}^{(k)} \gets \{ \texttt{boolean}(\mathbf{X}^{(k)}_j \neq \boldsymbol{\emptyset}) \mid j \in [1, D] \}$
                \State Store processed neural data and labeled neuron mask: $\mathbf{X}^{(k)}$ and $\mathbf{M}^{(k)}$
            \EndFor
            \State Save the processed dataset $\mathcal{D}_{\mathrm{processed, source}~i} := \{\left(\mathbf{X}^{(k)}, \mathbf{M}^{(k)}\right)\}_{k=1}^{N}$
        \EndFor
        \State Compile and store all processed datasets $\mathcal{D}_{\mathrm{processed, all\_sources}}$
    \end{algorithmic}
\end{algorithm}

\subsection{Connectome Graph Data}
We processed multiple open-source \textit{C. elegans} connectome datasets, each capturing the neural wiring diagram, including both chemical synapses and electrical synapse (also known as gap junctions). Table \ref{table:open_source_connectome_datasets} provides metadata for these datasets, which vary in annotation precision, formats, and the extent of the nervous system covered.

\subsubsection{Connectome Representation}
The connectome is represented as a directed graph \( \mathcal{G} = (\mathcal{V}, \mathcal{E}) \), where:

1. \textbf{Nodes (\( \mathcal{V} \))}: Represent individual neurons. 
\[
\mathcal{V} := \{\nu_i\}_{i=1}^{D}, \quad D := |\mathcal{V}| = 300.
\]
Each node \( \nu_i \) is associated with attributes, such as:
\begin{itemize}
    \item \textbf{Positional Coordinates}: Spatial location of the neuron in body-atlas coordinates \((x, y, z)\) \cite{skuhersky2022atlas}.
    \item \textbf{Neuron Class and Type}: Categories such as sensory, motor, or interneuron, and neuron-specific types (e.g., ADA, IL2, VD1).
    \item \textbf{Neurotransmitter}: Identity of neurotransmitters released by the neuron (e.g., acetylcholine, dopamine) \cite{hobert2016neuronal}.
    \item \textbf{Feature Vector (\( \mathbf{x}_i \))}: Optionally, numerical data (e.g. calcium fluorescence measurements over a fixed time window, gene expression embeddings)
\end{itemize}

2. \textbf{Edges (\( \mathcal{E} \))}: Represent synaptic and electrical connections between neurons.
\[
\mathcal{E} := \{e_{ij}\}_{k=1}^{E}, \quad E := |\mathcal{E}|.
\]
Each edge \( e_{ij} \) is associated with an attribute vector \( \mathbf{a}_{ij} \), which can include:
\begin{itemize}
    \item \( c \): Chemical synapse weight.
    \item \( g \): Gap junction weight.
    \item \( f \): Functional connectivity strength (optional, based on optogenetic perturbations \cite{randi2023neural}).
\end{itemize}

The neighbors of a node \( \nu_i \) are defined as \( \mathcal{N}(\nu_i) := \{\nu_j : e_{ij} \vee e_{ji} \in \mathcal{E}\} \), where \( \vee \) indicates "or", ignoring directionality.

Edges are included in \( \mathcal{E} \) if any of their attributes \( \mathbf{a}_{ij} \) are nonzero. For example, an edge may represent gap junctions (\( \mathbf{a}_{ij} = [0, g] \)) or chemical synapses (\( \mathbf{a}_{ij} = [c, 0] \)). In cases where functional connectivity is included, the graph may become fully connected (\( |\mathcal{E}| \approx D^2 \)) unless a significance threshold is applied to filter out weak functional connections. Chemical and electrical synapse weights are based on counts from EM annotations \cite{white1986structure, cook2019connectomes, witvliet2021connectomes} whereas the functional connectivity weights are based on signal propagation \cite{randi2023neural}.

\begin{figure}[!htbp]

\centering

\includegraphics[width=1.0\textwidth]{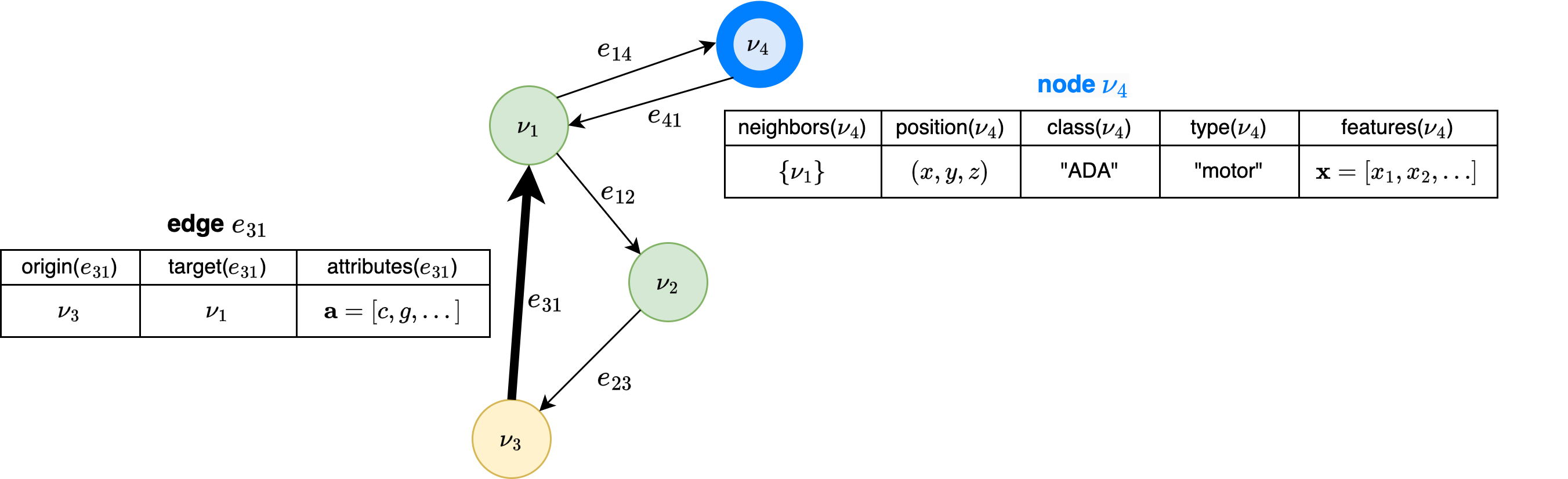}

\caption{\textbf{Graph-Based Representation of the Connectome.} Graph-based structure used for the \textit{C. elegans} connectome data. Nodes represent neurons, with node attributes (e.g., position, class) and optional features. Directed edges capture connections, with edge attributes (e.g., chemical and gap junction weights). }

\label{fig:connectome_example}

\end{figure}

Figure \ref{fig:connectome_example} illustrates this graph-based representation of the nervous system, highlighting node and edge attributes. Table \ref{table:graph_tensor_notation} summarizes the notation used in constructing and describing this graph structure.

\subsubsection{Data Extraction and Preprocessing}
The data extraction and preprocessing pipeline is implemented in the \texttt{ConnectomeBasePreprocessor} class, which standardizes neuron indices, processes edge attributes, and constructs the final graph tensor \texttt{Data} objects. Each source dataset is handled by a customized subclass since data extraction and preprocessing steps vary by dataset format:

\begin{enumerate}
    \item \textbf{Tabular Formats:}  
    Most datasets provide neuron pairs, connection types, and weights in a tabular structure. For example, the OpenWorm \cite{openworm2023} dataset specifies presynaptic and postsynaptic neurons, connection types (chemical or gap junction), and weights. See Table \ref{table:connectome_column_data_format} for an example.

    \item \textbf{Adjacency Matrix Formats:}  
    Some datasets use \( D \times D \) adjacency matrices, with entries representing synapse counts (\cite{cook2019connectomes}) or functional connectivity values (\cite{randi2023neural}).

    \item \textbf{Supplementing Attributes:}  
    Some desirable attributes about the connectome  (e.g., neuron positions, neurotransmitters, cell types, etc.) may not be available from the source datasets. We obtain these from  published research (e.g., neuron positions from \cite{skuhersky2022atlas}, neurotransmitters and cell types from \cite{hobert2016neuronal}) to supplement our graph structures.
\end{enumerate}

Custom classes within the preprocessing pipeline handle dataset-specific formats, ensuring consistent graph tensor outputs.

\subsubsection{Graph Tensor Format}

To facilitate downstream analyses and modeling, we standardized the connectome data into a graph tensor format compatible with the PyTorch Geometric graph neural network (GNN) libraries. Each dataset is transformed into a \texttt{torch\_geometric} \cite{fey2019torchgeometric} \texttt{Data} object with the following components:

\begin{itemize}
    \item \textbf{\texttt{edge\_index}}: Adjacency list encoding \( \mathcal{E} \).
    \item \textbf{\texttt{edge\_attr}}: Attribute vectors \( \mathbf{a}_{ij} = [c, g, f, \dots] \) for edges in \texttt{edge\_index}.
    \item \textbf{\texttt{x}}: Node feature matrix \( \mathbf{X} \in \mathbb{R}^{D \times L} \), initialized to zeros or populated with temporal slices of neural activity $\mathbf{X}^{k}[t:t+L]$ for some worm $k$. This data structure is the interface that allows integration of the neural activity data with connectivity data.
    \item \textbf{\texttt{kwargs}}: Metadata and node attributes including neuron coordinates, class, type, and neurotransmitter.
\end{itemize}

This tensor format ensures compatibility with downstream ML models and facilitates integration of functional (\( \mathbf{X} \)) and structural (\( \mathcal{E} \)) data. Table \ref{table:graph_tensor_notation} provides an overview of the standardized attributes.

\subsubsection{Consensus Connectome}

To address discrepancies in connection counts and weights across datasets, we construct a consensus connectome that aggregates synaptic data from multiple sources using a directional averaging strategy. Unlike the dataset-specific files used during preprocessing — which store information in graph network format with edge indices and attributes — the consensus connectome is a flat \texttt{.csv} table, not a graph. This tabular format is universal, self-explanatory, and easily readable by both humans and software tools. Each row corresponds to a directed pair of neurons, with clearly defined anatomical, connection, and metadata fields. This structure is especially suited for accessible downstream analysis, visualization, and integration, making it the preferred format for summarizing the overall state of connectivity across datasets.

For each ordered pair of neurons \( (i, j) \), we compute the consensus weights as follows:

\begin{itemize}
    \item \textbf{Gap Junctions} are symmetric. Their consensus weight \( g_{\text{consensus}} \) is computed as the mean of all non-zero gap weights observed in either direction across datasets:
    \[
    g_{\text{consensus}} = \operatorname{mean} \{g_{ij}^{(k)}\}
    \]
    where \( g_{ij}^{(k)} \) is the gap weight from neuron \( i \) to \( j \) in dataset \( k \). By definition, \( g_{ij}^{(k)} = g_{ji}^{(k)} \).

    \item \textbf{Chemical Synapses} are directional. Their consensus weight \( c_{\text{consensus}} \) is the mean of all non-zero chemical weights from neuron \( i \) to \( j \), excluding values from the functional connectivity dataset:
    \[
    c_{\text{consensus}} = \operatorname{mean} \{c_{ij}^{(k)} \}_{k \ne \text{funconn}}.
    \]

    \item \textbf{Functional Connectivity} is reported directly from the dedicated \texttt{funconn} dataset, if available. These values are not averaged and apply only in the directed case \( i \rightarrow j \).
\end{itemize}

To quantify the uncertainty in gap and chemical connection estimates, we compute a weighted standard deviation \( v \) that reflects both gap and chemical synapse variability:
\[
v_{\text{consensus}} = \lambda \cdot \operatorname{std}(\{c_{ij}^{(k)}\}_{k \ne \text{funconn}}) + (1 - \lambda) \cdot \operatorname{std}(\{g_{ij}^{(k)}\}_{k}),
\]
where \( \lambda = 0.9 \) is a subjective value chosen to weight the uncertainty from chemical synapses more due to their relative abundance compared to gap junctions.

\begin{figure}[!htbp]
\centering
\normalsize
\includegraphics[width=1.0\textwidth]{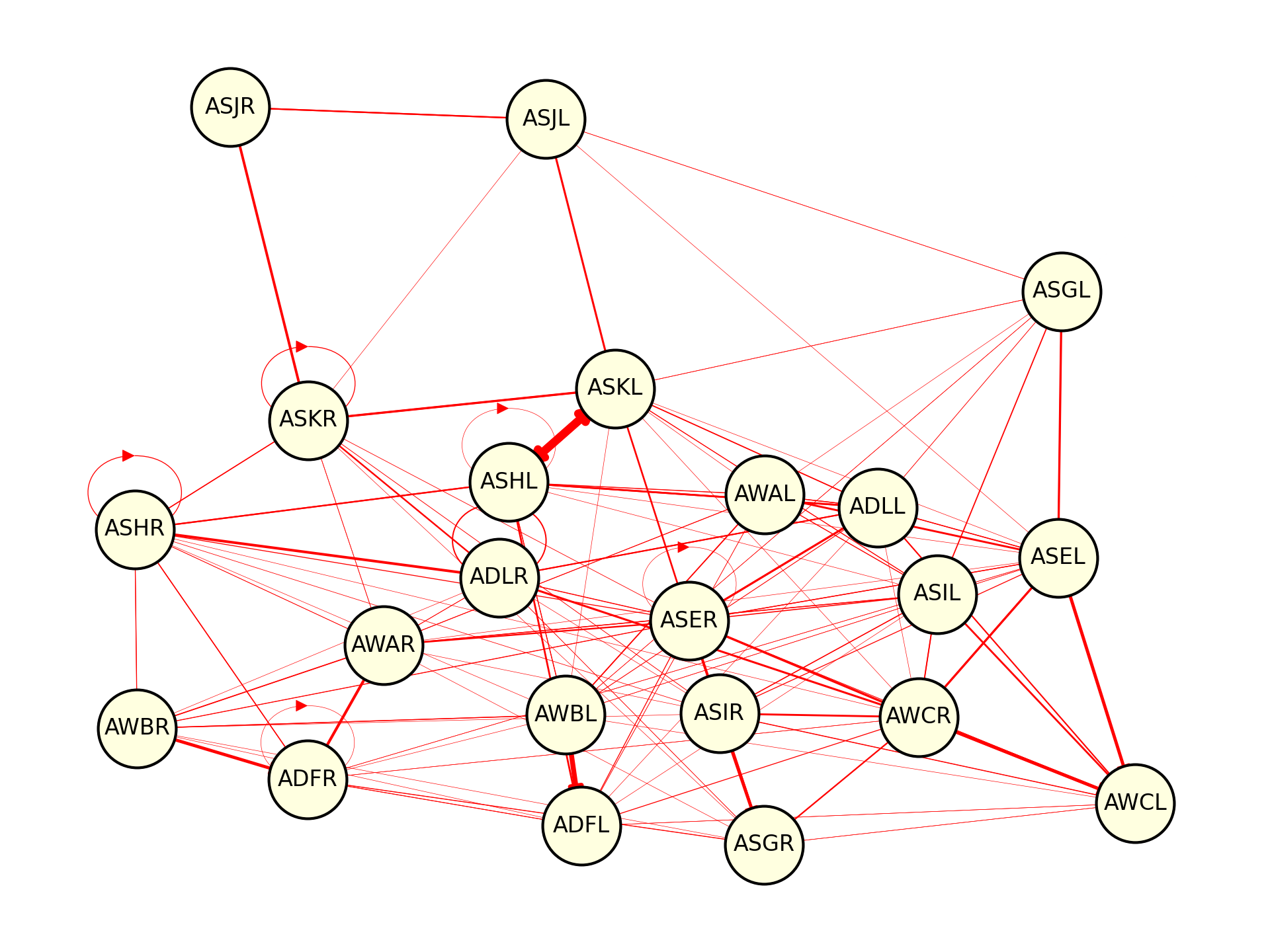}
Appendix Figure 1: \textbf{Consensus Connectome of Chemosensory Neurons.}. Visualization of the consensus connectome for a subset of the full network containing just the 22 amphid chemosensory neurons \cite{lin2023functional}. Nodes represent each neuron in 11 bilateral pairs; directed edges reflect connection strengths aggregated using the \texttt{mean} weight of each attribute across contributing datasets.
\label{fig:consensus_connectome}
\end{figure}

This consensus connectome includes all neuron pairs — including self-loops — and supports multiple output formats (e.g., trimmed vs. full, with or without functional weights). It provides a representative model of the stereotyped \textit{C. elegans} nervous system \cite{white1986structure}, while capturing cross-dataset variability that may reflect biological and technical noise. Figure~\ref{fig:consensus_connectome} visualizes the consensus connectivity among the 22 amphid chemosensory neurons \cite{lin2023functional}.


\section{Discussion}

In this paper we present standardized datasets for \textit{C. elegans} neural activity and connectomes. We have created and consolidated code blocks which can be used to load neural activity and connectome data from a variety of different source datasets with different file formats and subsequently preprocess this data. The code is composed of class definitions for each source dataset inheriting from a common \texttt{NeuralBasePreprocessor} class and a \texttt{ConnectomeBasePreprocessor} class along with helper functions which create class objects and execute class functions.

The homogenized neural and connectome datasets provide a unique opportunity to explore the relationship between neural dynamics and structural connectivity in a fully mapped nervous system. While our datasets provide a strong foundation for building models that link neural dynamics to structure, the limitations of both the neural activity data and the connectome must be taken into account when using these datasets. There are several important limitations and methodological choices that must be considered when interpreting the data and applying it in bio-computational models. We address most of these next.

One of the primary limitations of our neural activity dataset arises from the use of calcium fluorescence imaging as the method for measuring neural activity. Calcium imaging, while valuable for capturing broad patterns of activity across large populations of neurons, is an indirect measure of neural activity and introduces a low-pass filtering effect \cite{chen2013ultrasensitive}. Calcium signals lag behind the actual electrical activity due to the slower dynamics of calcium ion concentration changes compared to the fast voltage changes associated with action potentials. This delay can be particularly problematic for studies investigating fast synaptic interactions, as important rapid neural dynamics may be lost. Thus, while calcium imaging is useful for measuring slow, global brain states, it cannot fully capture the complexity of rapid information processing that may be critical in certain neural circuits \cite{tian2009imaging,wang2023voltage}. This highlights the need to complement calcium-based measurements with other techniques, such as those utilizing voltage-sensitive indicators, to gain a more comprehensive understanding of neural activity. However, calcium dynamics may be sufficient for the study of neural signaling in \textit{C. elegans} as most of its  neurons are nearly isopotential in the steady state \cite{goodman1998activecurrents}.

While the connectome of \textit{C. elegans} provides a complete structural blueprint of the nervous system, there are uncertainties and limitations inherent in structural connectomics. One major limitation is the variability in synaptic strengths between neurons, both across individuals and across the lifespan of a single individual \cite{witvliet2021connectomes}. The structural connectome is essentially a static snapshot, and we do not yet fully understand how synaptic strengths might fluctuate under different physiological conditions \cite{randi2023neural}. Moreover, the connectome data used in this study does not come from the same animals used for neural activity measurements, creating a mismatch between the dynamics and connectivity measurements. This creates both an opportunity and a limitation. The opportunity arises from the fact that any robust relationship between structure and function to be discovered from this data must generalize across different age and phenotype-matched animals. The limitation is that this may preclude the development of individual-specific models which are desirable for understanding learning and memory, for example. While having the \textit{C. elegans} connectome \cite{white1986structure, cook2019connectomes} provides a comprehensive structural map of synaptic connections, understanding the relationship between connectivity and dynamics still requires careful interpretation. Additionally, there is evidence for wireless connections between neurons \cite{randi2023neural} due to neuromodulation by way of neuropeptides which can bind to nearby and distant neuron GPCR sites triggering secondary messenger pathways \cite{beets2023system, ripoll2023neuropeptidergic}.

We opted not to apply smoothing to the neural activity traces for two principal reasons. First, it is likely that the original data creators have already performed a significant amount preprocessing on the signals extracted from their raw microscopy images, which often includes smoothing techniques to enhance signal quality. Introducing additional smoothing could risk over-processing the data, obscuring critical neural dynamics. Second, the calcium fluorescence signals utilized in our analysis are inherently low-pass filtered representations of the underlying neural activity, a characteristic of the imaging modality itself. This filtering effectively diminishes high-frequency components, which are vital for capturing rapid neural events. By refraining from further smoothing, we aim to preserve the integrity and richness of the original data, allowing for a more nuanced understanding of the biological phenomena under investigation.

Resampling was chosen as a strategy to make the data comparable across experiments with different temporal resolutions. We resampled the neural activity traces to a common time step of approximately $\Delta t \approx 0.333$ seconds, which was chosen as a balance between temporal resolution and minimizing data loss due to downsampling from experiments with higher frame rates. However, this decision introduces its own limitations. There was a tradeoff between attenuating high-frequency information from datasets with higher sampling rates versus incorrectly hallucinating data from datasets with lower sampling frequencies. For experiments that originally had much higher temporal resolution, downsampling risks losing potentially valuable high-frequency dynamics. Conversely, for lower-resolution datasets, resampling can artificially inflate the temporal resolution without actually adding more meaningful information \cite{lin2023functional}. This was a necessary compromise to allow for cross-study comparisons but should be considered when interpreting the results.

One important clarification about our work needs to be made regarding our use of $300$ for the number of neurons in \textit{C. elegans}, whereas readers may be more familiar the number $302$ from prior literature. We drop the bilateral pair of canal-associated neurons (CAN) from our datasets because they completely lack chemical synapses and are now widely classified as `end-organs' rather than neurons \cite{stefanakis2015panneuronal, ripoll2023neuropeptidergic}. The absence of synaptic polarity information, as mentioned earlier, limits the ability of models to fully capture the functional consequences of the structural connections. Moreover, the use of calcium measurements might bias neural activity models towards capturing slow, large-scale patterns at the expense of finer, faster neural interactions. Researchers should be mindful of these constraints when developing models and consider integrating additional multimodal datasets to build more comprehensive models \cite{haspel2023reverseengineer}.

Despite these limitations, the \textit{C. elegans} neural dataset represents a valuable resource for computational research, particularly in developing biologically grounded models of neural dynamics. For instance, graph neural networks (GNNs) \cite{velickovic2023everything} are well-suited to handle the structural connectivity data, while time-series models like structured state space models (SSMs) \cite{gu2021statespaces} can be applied to the neural activity data to capture temporal dependencies. These models could inform the development of foundation models that generalize to larger and more complex systems. Moreover, the datasets' open availability facilitates collaborative efforts in both neuroscience and AI, encouraging the development of new techniques and models that can integrate structure and function in novel ways.

In conclusion, while our homogenized datasets offer significant opportunities for advancing our understanding of small neural systems, its limitations should be carefully considered in both biological and computational contexts. Calcium imaging, noisy connectomics, and the standardization choices we made each come with trade-offs that impact how the data can be used. As neural imaging technologies continue to improve and new datasets are integrated, we expect that this dataset will serve as a foundation for building more complete models of neural function, bridging the gap between simple nervous systems like \textit{C. elegans} and the more complex neural architectures found in larger organisms.

\section{Conclusion}

We present a standardization protocol implemented in a \href{https://github.com/qsimeon/worm-data-preprocess}{code repository} for the preprocessing and integration of 12 \textit{C. elegans} calcium fluorescence datasets and 4 primary connectome publications \cite{white1986structure,cook2019connectomes,witvliet2021connectomes,randi2023neural}. The code repository contains preprocessing classes that load and preprocess the different source datasets, handling both calcium fluorescence and connectome data. Each dataset has a class implementation that inherits from a respective parent class, which has been adapted to accommodate varying raw data formats. Once loaded, the calcium fluorescence data is further masked, normalized, resampled, and smoothed. 

This unified dataset offers an accessible and standardized resource, streamlining access to data recorded under diverse experimental conditions and formats. The datasets are available for public access on Hugging Face: (1) \href{https://huggingface.co/datasets/qsimeon/celegans_neural_data}{neural data}, and (2) \href{https://huggingface.co/datasets/qsimeon/celegans_connectome_data}{connectome data}. Additionally, the full code repository containing the preprocessing code can be accessed on \href{https://github.com/qsimeon/worm-data-preprocess}{GitHub}.

Such standardization facilitates the development of models for the entire \textit{C. elegans} nervous system, allowing researchers to study the interplay between structure, via the connectome, and function, via calcium fluorescence data. Through this integration, we hope to provide further insights into how neural function emerges from structural connectivity and how these dynamics might generalize to more complex nervous systems.

The open-source nature of this resource aims to further contribute to both neuroscience and AI research, accelerating efforts to model not only \textit{C. elegans} but also organisms with larger and more complex nervous systems. By releasing this dataset and accompanying code, we hope to advance the growing initiative toward neural structure-function modeling and the discovery of principles of neural computation across species.

\clearpage

\nocite{*} 
\bibliographystyle{plainnat}
\bibliography{references}
\newpage

\section{Appendix}

\begin{table}[!htbp]
\centering
\normalsize 
Appendix Table 1: \textbf{\textit{C. elegans} Neurons Categorized by Type.} Neurons in \textit{C. elegans} are categorized by their type: motor, inter, sensory, and pharynx neurons. Neuron names are listed under their respective types.
\scriptsize 
\renewcommand{\arraystretch}{1.2} 
\begin{tabularx}{\textwidth}{|l|X|}
\hline
\textbf{Neuron Type} & \textbf{Neuron Names} \\
\hline
Motor Neurons & AS1, AS2, AS3, AS4, AS5, AS6, AS7, AS8, AS9, AS10, AS11, AVL, DA1, DA2, DA3, DA4, DA5, DA6, DA7, DA8, DA9, DB1, DB2, DB3, DB4, DB5, DB6, DB7, DD1, DD2, DD3, DD4, DD5, DD6, DVB, HSNL, HSNR, PDA, PDB, RIML, RIMR, RMDDL, RMDDR, RMDL, RMDR, RMDVL, RMDVR, RMED, RMEL, RMER, RMEV, RMFL, RMFR, RMHL, RMHR, SABD, SABVL, SABVR, SIADL, SIADR, SIAVL, SIAVR, SIBDL, SIBDR, SIBVL, SIBVR, SMBDL, SMBDR, SMBVL, SMBVR, SMDDL, SMDDR, SMDVL, SMDVR, VA1, VA2, VA3, VA4, VA5, VA6, VA7, VA8, VA9, VA10, VA11, VA12, VB1, VB2, VB3, VB4, VB5, VB6, VB7, VB8, VB9, VB10, VB11, VC1, VC2, VC3, VC4, VC5, VC6, VD1, VD2, VD3, VD4, VD5, VD6, VD7, VD8, VD9, VD10, VD11, VD12, VD13 \\
\hline
Inter Neurons & ADAL, ADAR, AIAL, AIAR, AIBL, AIBR, AIML, AIMR, AINL, AINR, AIYL, AIYR, AIZL, AIZR, AUAL, AUAR, AVAL, AVAR, AVBL, AVBR, AVDL, AVDR, AVEL, AVER, AVFL, AVFR, AVG, AVHL, AVHR, AVJL, AVJR, AVKL, AVKR, BDUL, BDUR, DVA, DVC, LUAL, LUAR, PVCL, PVCR, PVNL, PVNR, PVPL, PVPR, PVQL, PVQR, PVR, PVT, PVWL, PVWR, RIAL, RIAR, RIBL, RIBR, RICL, RICR, RID, RIFL, RIFR, RIH, RIPL, RIPR, RIR, RIS, RIVL, RIVR \\
\hline
Sensory Neurons & ADEL, ADER, ADFL, ADFR, ADLL, ADLR, AFDL, AFDR, ALA, ALML, ALMR, ALNL, ALNR, AQR, ASEL, ASER, ASGL, ASGR, ASHL, ASHR, ASIL, ASIR, ASJL, ASJR, ASKL, ASKR, AVM, AWAL, AWAR, AWBL, AWBR, AWCL, AWCR, BAGL, BAGR, CEPDL, CEPDR, CEPVL, CEPVR, FLPL, FLPR, IL1DL, IL1DR, IL1L, IL1R, IL1VL, IL1VR, IL2DL, IL2DR, IL2L, IL2R, IL2VL, IL2VR, OLLL, OLLR, OLQDL, OLQDR, OLQVL, OLQVR, PDEL, PDER, PHAL, PHAR, PHBL, PHBR, PHCL, PHCR, PLML, PLMR, PLNL, PLNR, PQR, PVDL, PVDR, PVM \\
\hline
Pharynx Neurons & I1L, I1R, I2L, I2R, I3, I4, I5, I6, M1, M2L, M2R, M3L, M3R, M4, M5, MCL, MCR, MI, NSML, NSMR \\
\hline
\end{tabularx}
\label{table:celegans_neuron_types}
\end{table}

\begin{table}[!htbp]
\normalsize
Appendix Table 2: \textbf{Extended metadata for Table \ref{table:open_source_neural_datasets}}. A link to download each neural activity source dataset as well as regular expressions matching the files from which the raw data was extracted is included.
\hfill
\centering
\scriptsize
\renewcommand{\arraystretch}{1.2}
\begin{tabularx}{\columnwidth}{|X|X|X|X|}
\hline
\textbf{Source Dataset Index} & \textbf{Source Dataset Name} & \textbf{Database Link} & \textbf{Data Files} \\
\hline
1 & Kato2015 \cite{kato2015global} & \href{https://osf.io/2395t/}{OSF:2395t} & WT\_\*Stim.mat \\
2 & Nichols2017 \cite{nichols2017brain} & \href{https://osf.io/kbf38/}{OSF:kbf38} & \*let.mat \\
3 & Skora2018 \cite{skora2018energy} & \href{https://osf.io/za3gt/}{OSF:za3gt} & WT\_\*.mat \\
4 & Kaplan2020 \cite{kaplan2020nested} & \href{https://osf.io/9nfhz/}{OSF:9nfhz} & Neuron2019\_Data\_*.mat \\
5 & Nejatbakhsh2020 \cite{nejatbakhsh2020} & \href{https://dandiarchive.org/dandiset/000541?pos=1}{DANDI:000541} & sub-YYYYMMDD-wormidx \\
6 & Yemini2021 \cite{yemini2021neuropal} & \href{https://zenodo.org/record/3906530}{Zenodo:3906530} & \*\_Activity\_OH\*.mat \\
7 & Uzel2022 \cite{uzel2022hubneurons} & \href{https://osf.io/3vkxn/}{OSF:3vkxn} & Uzel\_WT.mat \\
8 & Dag2023 \cite{dag2023serotonergic} & \href{https://tinyurl.com/githubDag2023}{GitHub:Dag2023} & data/swf702\_with\_id/*.h5 \\
9 & Atanas2023 \cite{atanas2023brainwide} & \href{https://wormwideweb.org}{wormwideweb.org} & YYYY-MM-DD-\*.json/h5 \\
10 & Leifer2023 \cite{randi2023neural} & \href{https://tinyurl.com/driveLeifer2023}{Drive:Leifer2023} & exported\_data.tar.gz \\
11 & Lin2023 \cite{lin2023functional} & \href{https://tinyurl.com/dropboxLin2023}{Dropbox:Lin2023} & run\*\_prfrd\_data\*.mat \\
12 & Venkatachalam2024 \cite{venkatachalam2024website} & \href{https://chemosensory-data.worm.world}{worm.world} & 2022\*\_herm\_\*.zip \\
\hline
\end{tabularx}
\label{table:extended_metadata_table1}
\end{table}

\newpage 

\begin{table}[!htbp]
\normalsize
Appendix Table 3: \textbf{Mathematical Notation Used in Neural Data Preprocessing.} Summary of the key symbols and their descriptions used in preprocessing \textit{C. elegans} neural activity datasets. These notations represent worm-specific indices, neural activity structures, temporal properties, neuron counts, and preprocessing parameters. Additional details from the methods section are included to aid readers unfamiliar with our mathematical notation.
\hfill
\centering
\scriptsize
\renewcommand{\arraystretch}{1.3} 
\begin{tabularx}{\textwidth}{|c|X|}
\hline
\textbf{Symbol} & \textbf{Description} \\
\hline
$k$ & Index used for worm in a single dataset \\
$t, t'$ & Original and resampled time indices for the neural activity data \\
$T_k$ & Number of time points recorded for worm $k$ \\
$D$ & Total number of canonical neurons in \textit{C. elegans} ($D=300$ since we exclude the CAN cells \cite{stefanakis2015panneuronal, ripoll2023neuropeptidergic}) \\
$i, j$ & Indices used for the neuron dimension. $i,j \in \{1, \dots, D\}$ \\
$\mathbf{X}^{(k)}$ & Neural activity data matrix for a single worm $k$. $\mathbf{X}^{(k)} \in \mathbb{R}^{T_k \times D}$ \\
$\mathbf{X}^{(k)}[t]$ & Neural activity vector at time $t$ for all neurons, $\mathbf{X}^{(k)}[t] \in \mathbb{R}^D$ \\
$\mathbf{X}^{(k)}_{j}$ & Neural activity time series for neuron $j$ in worm $k$. $\mathbf{X}^{(k)}_{j} \in \mathbb{R}^{T_k}$ \\
$\boldsymbol{\mu}, \boldsymbol{\sigma}$ & Mean and standard deviation over time of neural activity data for each neuron \\
$\mathbf{M}^{(k)}$ & Binary mask indicating labeled neurons for worm $k$. $\mathbf{M}^{(k)} \in \mathbb{R}^D$ \\
$\alpha$ & Smoothing hyperparameter for exponentially weighted moving average (default: 1) \\
$\Delta t$ & Constant resampling time interval for calcium fluorescence sampling (default: 0.333 seconds) \\
$\Delta t_{\text{original}}$ & Original and potentially variable time resolution of the dataset before resampling \\
\hline
\end{tabularx}
\label{table:neural_notation_table}
\end{table}

\begin{table}[!htbp]
\normalsize
Appendix Table 4: \textbf{Mathematical Notation for Connectome Graph Representation.} Key symbols and their descriptions used in the graph-based representation and tensor format.
\hfill
\centering
\scriptsize
\renewcommand{\arraystretch}{1.3}
\begin{tabularx}{\textwidth}{|c|X|}
\hline
\textbf{Symbol} & \textbf{Description} \\
\hline
\( \mathcal{G} = (\mathcal{V}, \mathcal{E}) \) & Directed graph representation of the connectome, where \( \mathcal{V} \) is the set of nodes (neurons) and \( \mathcal{E} \) is the set of directed edges (connections). \\
\( \mathcal{V}, D \) & Set of neurons (\( \mathcal{V} \)) and total count (\( D := 300 \)), corresponding to the neurons in the \textit{C. elegans} nervous system. \\
\( \mathcal{E}, E \) & Set of directed edges (\( \mathcal{E} \)) and total count (\( E := |\mathcal{E}| \)), where edges represent chemical synapses, gap junctions, or other connections. \\
\( \nu_i \) & Node \( i \), representing the \( i \)-th neuron in the graph. \\
\( \mathbf{x}_i \) & Feature vector associated with neuron \( \nu_i \), optionally containing temporal data (e.g., calcium fluorescence). \\
\( \mathcal{N}(\nu_i) \) & Set of neighbors of neuron \( \nu_i \), defined as \( \{\nu_j : e_{ij} \vee e_{ji} \in \mathcal{E}\} \). \\
\( e_{ij} \) & Directed edge from neuron \( \nu_i \) to \( \nu_j \), representing a connection. \\
\( \mathbf{a}_{ij} \) & Attribute vector for edge \( e_{ij} \), containing attributes such as \( c \), \( g \), and \( f \). \\
\( c \) & Chemical synapse weight between neurons \( \nu_i \) and \( \nu_j \). If no synapse exists, \( c = 0 \). \\
\( g \) & Gap junction (electrical synapse) weight between neurons \( \nu_i \) and \( \nu_j \). If no gap junction exists, \( g = 0 \). \\
\( f \) & Functional connectivity strength between neurons \( \nu_i \) and \( \nu_j \), derived from optogenetic perturbations \cite{randi2023neural}. Optional in the attribute vector \( \mathbf{a}_{ij} \). \\
\( x, y, z \) & Positional coordinates of neuron \( \nu_i \) in body-atlas space. \\
\texttt{edge\_index} & Matrix encoding the adjacency structure of \( \mathcal{E} \). Each column contains a pair of indices \( (i, j) \) for an edge \( e_{ij} \). \\
\texttt{edge\_attr} & Tensor of edge attributes, where each row corresponds to the attribute vector \( \mathbf{a}_{ij} \) for an edge \( e_{ij} \). \\
\texttt{x} & Node feature matrix \( \mathbf{X} \in \mathbb{R}^{D \times L} \), where each row represents the feature vector \( \mathbf{x}_i \) of a neuron. \\
\texttt{kwargs} & Dictionary containing metadata for neurons, such as class, type, neurotransmitter identity, and coordinates. \\
\hline
\end{tabularx}
\label{table:graph_tensor_notation}
\end{table}

\begin{table}[!htbp]
\normalsize
Appendix Table 5: \textbf{Metadata for Open-Source Connectome Datasets.} Metadata for open-source \textit{C. elegans} connectome datasets standardized into the graph tensor format. The table includes the graph tensor file and the total number of edges (synaptic and electrical connections).
\hfill
\centering
\scriptsize
\renewcommand{\arraystretch}{1.3}
\begin{tabularx}{\textwidth}{|l|c|c|X|}
\hline
\textbf{Source Dataset Name} & \textbf{Number of Nodes} & \textbf{Number of Edges} & \textbf{Created Graph Tensor File} \\
\hline
Cook2019 \cite{cook2019connectomes} & 236 & 4085 & \texttt{graph\_cook2019.pt} \\
White1986 (whole) \cite{white1986structure} & 299 & 3145 & \texttt{graph\_white1986\_whole.pt} \\
White1986 (n2u) \cite{white1986structure} & 180 & 1848 & \texttt{graph\_white1986\_n2u.pt} \\
White1986 (jsh) \cite{white1986structure} & 178 & 1783 & \texttt{graph\_tensors\_white1986\_jsh.pt} \\
White1986 (jse) \cite{white1986structure} & 34 & 94 & \texttt{graph\_tensors\_white1986\_jse.pt} \\
Witvliet2020 (7) \cite{witvliet2021connectomes} & 180 & 2364 & \texttt{graph\_tensors\_witvliet2020\_7.pt} \\
Witvliet2020 (8) \cite{witvliet2021connectomes} & 180 & 2371 & \texttt{graph\_tensors\_witvliet2020\_8.pt} \\
OpenWorm2023 \cite{openworm2023} & 299 & 3142 & \texttt{graph\_tensors\_openworm.pt} \\
Randi2023 \cite{randi2023neural} & 167 & 11175 & \texttt{graph\_tensors\_funconn.pt} \\
Chklovskii2011 \cite{varshney2011structural} & 215 & 2079 & \texttt{graph\_tensors\_chklovskii.pt} \\
\hline
\end{tabularx}
\label{table:open_source_connectome_datasets}
\end{table}

\begin{table}[!htbp]
Appendix Table 6: \textbf{Column Data Format for OpenWorm Connectome Dataset.} Example of tabular format from the OpenWorm \cite{openworm2023} connectome source file, listing origin and target neurons, connection type, and weights.
\centering
\scriptsize
\begin{tabularx}{\columnwidth}{|X|X|X|X|X|X|}
\hline
\textbf{Origin} & \textbf{Target} & \textbf{Type} & \textbf{Number of Connections} & \textbf{Neurotransmitter} \\
\hline
 ADAL & ADAR & GapJunction & 1 & Generic\_GJ \\
 ADAL & ADFL & GapJunction & 1 & Generic\_GJ \\
 ADAL & AIBL & Send & 1 & Glutamate \\
 ADAL & AIBR	& Send & 2 & Glutamate \\
 ADAL & ASHL	& GapJunction & 1 & Generic\_GJ\\
 ... & ... & ... & ... & ... \\
 VC1	& VC2 & Send & 3 & Serotonin\_Acetylcholine \\
 VC1	& VC2 & GapJunction & 6 & Generic\_GJ \\ 
 VC1	& VC3 & Send & 1 & Serotonin\_Acetylcholine \\
 VC1	& VC3 & GapJunction & 2 & Generic\_GJ \\
 VC1	& VD1 & GapJunction	& 1	& Generic\_GJ \\
\hline
\end{tabularx}
\label{table:connectome_column_data_format}
\end{table}

\newpage 

\end{document}